\documentclass[preprint,review,12pt]{elsarticle}
\usepackage{scrhack} 
\usepackage{listings}
\usepackage{lstautogobble}

\AtBeginDocument{%
	}

\begin{document}
	
	\title{Contiguous Storage of Grid Data for Heterogeneous Computing}
	
\author{Fan Gu}
\ead{ge86gur@mytum.de}
\author{Xiangyu Hu \corref{mycorrespondingauthor}}
\cortext[mycorrespondingauthor]{Corresponding author.}
\ead{xiangyu.hu@tum.de}
\address{Technical University of Munich, Garching 85748, Germany}

	\begin{abstract}
		Structured Cartesian grids are a fundamental component in numerical simulations. 
		Although these grids facilitate straightforward discretization schemes, 
		their na\"{i}ve use in sparse domains leads to excessive memory overhead and inefficient computation. 
		Existing frameworks address are primarily optimized for CPU execution 
		and exhibit performance bottlenecks on GPU architectures due to limited parallelism 
		and high memory access latency. 
		This work presents a redesigned storage architecture optimized for GPU compatibility 
		and efficient execution across heterogeneous platforms. 
		By abstracting low-level GPU-specific details and adopting a unified programming model based on SYCL, 
		the proposed data structure enables seamless integration across host and device environments. 
		This architecture simplifies GPU programming for end-users while improving scalability 
		and portability in sparse-grid and gird-particle coupling numerical simulations.
	\end{abstract}
	\begin{figure}[htb!]
		\centering
		\includegraphics[width=\textwidth]{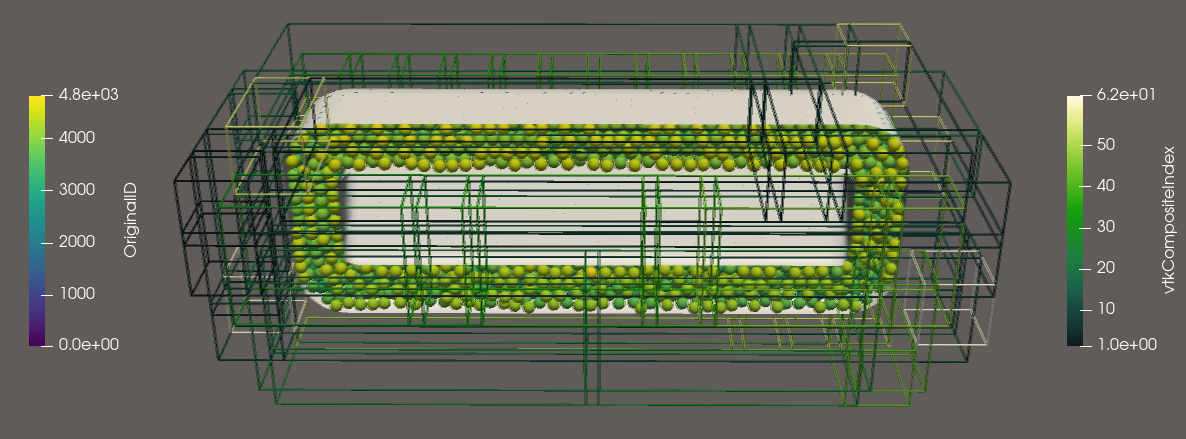}
		\caption{Particle generation from an extruded prism.
			The gray surface indicates the clipped zero level set, the framed boxes indicate the core data package blocks used to describe the surface and the small spheres indicate the SPH particles generated and physically relaxed for later numerical simulation.}
		\label{fig:teaser}
	\end{figure}
	\maketitle
	
	\section{Introduction}
	Many numerical methods in scientific computing, computer graphics, and computational physics rely on structured Cartesian grids to discretize scalar and vector fields. These grids are widely used in level set methods\cite{osher2001level}, computational fluid dynamics (CFD) 
	solvers \cite{enright2002hybrid, han2014adaptive}, 
	and volumetric data processing pipelines \cite{museth2013vdb}. 
	While Cartesian grids offer simplicity and regularity for finite difference and finite volume methods, large-scale simulations often involve highly sparse domains where only a small fraction of the grid is actively used at any time.
	
	Naively storing and updating such sparse data results in excessive memory consumption 
	and suboptimal memory access patterns, 
	particularly on modern hardware architectures 
	with hierarchical memory system and parallel processing constraints. 
	To address this, data structures such as OpenVDB \cite{museth2013vdb} 
	and SPGgrid \cite{setaluri2014spgrid} have been developed to reduce memory overhead 
	and improve computational performance. 
	These frameworks leverage hierarchical spatial partitioning or page-based layouts to exploit sparsity. 
	However, they are primarily optimized for CPU-based execution, 
	and often suffer from performance degradation when ported to Graphic Processing Unit (GPU) architectures. 
	Key challenges include high random-access latency and limited concurrency during data update.
	
	GPU has become prevalent in accelerating a wide range of numerical simulations 
	due to their high parallel processing capability. 
	Despite the high computing power provided, 
	leveraging GPU typically requires the adoption of specialized programming models, 
	such as CUDA(NVIDIA), HIP(AMD), SYCL, and others, 
	which presents a steep learning curve for end-users.
	
	In our prior implementations in the open-source SPH (Smoothed Particle Hydrodynamics) 
	multi-physics library SPHinXsys \cite{han2014adaptive, zhang2021sphinxsys}, 
	memory allocation for activated grid regions was managed via dynamic memory pools. 
	While this approach yields efficient memory usage, 
	the resulting data structures are not inherently compatible with GPU execution. 
	In this work, we present a redesigned storage architecture aimed at improving computational efficiency and enabling GPU compatibility. 
	The new design is optimized for unified usage across both host and device platforms, 
	abstracting underlying implementation details from the user.
	
	This work introduces an optimized data structure tailored for GPU execution, 
	along with a newly developed computing kernels. 
	The key contributions are:
	\begin{enumerate}
		\item Abstraction of GPU-specific and SYCL-specific implementation details, thereby simplifying usage for end-users.
		\item Unified codebase enabling seamless execution on both CPU and GPU platforms through a single code logic.
		\item Heterogeneous computing is achieved for both sparse-grid and grid-particle coupling numerical simulations.
	\end{enumerate}
	
	The proposed method incorporates computing kernels, data structures and execution strategies. 
	The computing kernel orchestrates the computational process, 
	while data structure manage storage and synchronization of data across CPU and GPU. 
	Upon execution, the computing kernel automatically retrieves the appropriate host 
	or device variable based on the selected execution strategy, 
	thus ensuring consistency and portability.
	
	\section{Literature Overview}
	OpenVDB \cite{museth2013vdb} employs a shallow B+ trees to store data efficiently, 
	using various strategies to accelerate both sequential and stencil-based data access. 
	By dynamically organizing data with a hierarchical B+ tree, 
	it achieves memory efficiency by storing only activated cells within the leaf nodes. 
	A software cache is utilized to optimize traversal, 
	exploiting spatial locality between successive accesses.
	
	Despite its compact storage model, 
	and being widely adopted in the visual effects industry,
	scientific visualization and game development,
	the dynamic nature of OpenVDB's memory allocation presents challenges for GPU acceleration. 
	This is particularly evident when operating on high-resolution levels, 
	where the cost of frequent memory allocation and pointer chasing inhibits parallel execution efficiency. 
	While dynamic topology updates are advantageous for evolving level set geometries, 
	the tree-based storage of leaf nodes introduces access overhead, especially for stencil operations, 
	as high-speed random-access of the activated data require 
	both fast index direction and contiguous memory.

	In response to OpenVDB's limitations of CPU-only workflows, 
	the NanoVDB method \cite{museth2021nanovdb} has emerged to support GPU execution. 
	Tuned for visualization, it provides a read-only, flattened representation of OpenVDB grids 
	that can be transferred to the GPU. 
	Topological changes and data modifications are performed exclusively on the host, 
	after which the updated grid is converted into the NanoVDB format and reloaded onto the device for use. 
	While this enables limited GPU utilization, 
	it restricts dynamic interactions and update concurrency on the device.
	
	To optimize memory access patterns in OpenVDB,
	the SPGrid method \cite{setaluri2014spgrid} 
	organizes data within specially allocated arrays and exploits hardware characteristics. 
	It presents a hybrid model that combines the structural regularity of dense grids 
	with the memory efficiency of sparse representations. 
	In contrast to pointer-based structures, 
	SPGrid utilizes a compact, 
	linearized memory layout to represent active regions within a sparse, uniform grid.
	
	SPGrid reserves a virtual memory address space proportional to the full mesh resolution, 
	while allocating physical memory only for cells that are actively used. This strategy supports efficient index translation through the use of Translation Lookaside Buffer(TLB) 
	and preserves spatial locality, 
	yielding performance comparable to dense grids while using significantly less memory. 
	However, the total mesh size that SPGrid can accommodate remains constrained 
	by the available main memory, 
	as it relies on virtual memory addressing schemes managed at the system level.
	Furthermore, although SPGrid demonstrates high performance on CPUs—particularly 
	for workloads demanding structured sparsity, cache locality, 
	and high arithmetic intensity, its design is not inherently compatible with GPU execution. 
	This is primarily due to its dependence on CPU-style memory addressing, 
	pointer arithmetic, and indirect accesses, all of which limit the efficiency of GPU parallelization.
	\section{Design Objectives and Main Components}
	The present objective of using sparse-grid storage in SPHinXsys is two folded.
	One is to represent the body surfaces with a level set field, 
	similar to the application of visualization,
	which is used to generate SPH particles for numerical simulation. 
	The other is level-set based operations, 
	which are both computation and memory intensive, 
	including small-feature cleaning, 
	water-tight-surface ensuring and 
	convolution integrals with SPH smoothing kernel \cite{yu2023level}.
	Note that these operations are essential for dynamically reorganizing 
	the SPH particles to be body-fitted and well distributed 
	for stable and accurate numerical simulations.
	
	Therefore, the present method is composed of three primary components as 
	sparse-grid (narrow-band) storage, access and computation, respectively. 
	While the first component is managed by the \texttt{MeshWithGridDataPackage} class, 
	as shown in Lst. \ref{lst:mesh_class},
	the second provides the functions for computing stencils 
	of interpolations and differentials based on the access function \texttt{NeighbourIndexShift}.
	Two patterns of computation, 
	i.e. \texttt{ALLMeshDynamics} and \texttt{MeshPackageDynamics},
	are provided to transverse the entire Cartesian grid 
	and activated cells, respectively, 
	with execution policies (e.g., sequential, 
	or parallel executions on host and device) to be applied.
	\section{\texttt{MeshWithGridDataPackages}}
\begin{figure}[htb!]
		\centering
		\includegraphics[width=0.75\linewidth]{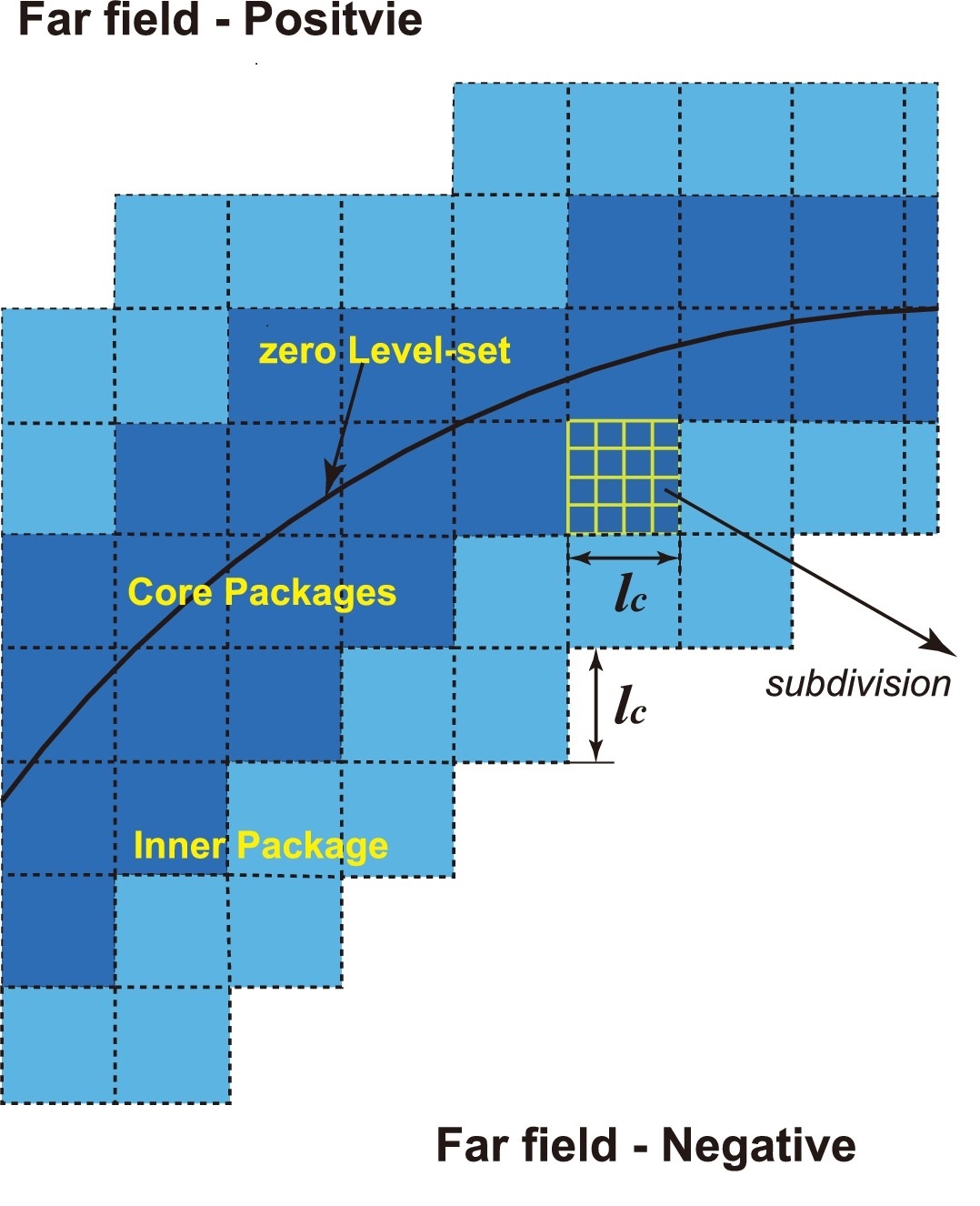}
		\caption{Level-set field with narrow band data-package storage: ‘inner packages’(light blue cells) and ‘core packages’(dark blue cells) around the surface. 
			$l_{c}$ indicates the coarse cell size of the background mesh. 
			The far field negative and positive represent the far regions within and outside the body surface.}
		\label{fig:data-package}
	\end{figure}

	The \texttt{MeshWithGridDataPackages} class defines a coarse (background) mesh 
	in which a portion of cells are activated (cut by the zero-level-set) 
	and used to store refined level-set values  
	in the form of block-data packages with a prescribed subdivision size (default by 4),
	as shown in Figs. \ref{fig:teaser} and  \ref{fig:data-package}. 
	There are two types of data-packages: 
	one is the core ones for describing body surface 
	and the other inner ones used for level-set operations.    
	
	By making use of the \texttt{DiscreteVariable} type, 
	which manages 1D array of particle-wise data in SPHinXsys, 
	we defines three types of variables for the present method. 
	One is \texttt{BKGMeshVariable} 
	with the size of total number background mesh cells 
	for indicating the activation-status of the cells 
	or the index of data package if the cell is activated. 
	One is \texttt{MeshVariable} for level-set data-packages 
	with the size of total activated cells. 
	The other is \texttt{MetaVariable} with the same size of \texttt{MeshVariable} 
	but using simple data to define the type, cell location and topology (neighborhood) 
	of the data packages. 
	Note that, due to the coarse background mesh and the sparse activation,
	the amount of memory usage for \texttt{BKGMeshVariable} and 
	\texttt{MetaVariable} 
	are negligible compared to that for \texttt{MeshVariable}.
	
	{
		\scriptsize
		\begin{lstlisting}[caption={Simplified structure of \texttt{MeshWithGridDataPackages}}, label={lst:mesh_class}, language=C++]
		template <int PKG_SIZE>
		class MeshWithGridDataPackages
		{
		//spacing of data in data package
		const Real data_spacing_;
		
		DiscreteVariable<CellNeighborhood> cell_neighborhood_;
		BKGMeshVariable<size_t> cell_package_index_;
		DiscreteVariable<std::pair<size_t, int>> meta_data_cell_;
		
		MeshVariableAssemble all_mesh_variables_;
		}
		\end{lstlisting} 
	}
	
	With registration, 
	each \texttt{MeshVariable} is assigned with a name identifier 
	and inserted into the appropriate type-specific data assembly. 
	Retrieval operations are subsequently performed using both the variable's type 
	and its assigned name. 
	Note that a \texttt{MeshVariable} must be registered prior to any usage.
	A generic operation interface is provided 
	to enable type-agnostic operations across all registered mesh variables. 
	The interface currently supports two primary operations:
	
	\begin{enumerate}
		\item Reallocation: Upon requested, the memory for data packages is reallocated either on host or device.
		\item Synchronization: Following computations on the device side, 
		the data of mesh variables specifically registered for file output is synchronized back to the host. 
	\end{enumerate}
	
	Note that, 
	these operation interface guarantees consistency 
	across all supported data types.
	\section{Array Storage and Access}
	Compared to the usage of memory pool for dynamic allocation, 
	which is fundamentally incompatible with GPU architectures,
	in previous SPHinXsys implementation, 
	the present method allocates 1D array, i.e. contiguous memory 
	for all activated data packages (actual data of all above defined variables),
	improving memory utilization and supporting more efficient GPU execution
	due to their structure-of-arrays (SoA) layout. 
	Note that, the first two entries of the array are reserved for the two packages 
	representing negative and positive far fields, as shown in Fig. \ref{fig:data-package},
	the other activated data packages is allocated after them.
	\subsection{Indexing}
	For easy direct access of the mesh variables, a indexing system is employed, 
	comprising three types: package, cell and data indexes.
	
	\begin{enumerate}
		\item {\it Package index} is a unique identifier assigned to each activated cell of the mesh. 
		It is used to index all relevant mesh variable data packages. 
		That is, regardless of the specific data type being accessed 
		(e.g., scalar and vector fields), 
		the same package index will be used to retrieve data corresponding to the same activated cell.
		\item {\it Cell index} is referring to the linear cell location of an activated cell of the mesh. 
		This index plays a critical role in mapping physical space to the data packages 
		and is essential for identifying activated cells during initialization and spatial queries.
		\item {\it Data index} represents the local index of 
		a data point within a data package, ranging from 0 to 3 by default for each dimension. 
		It is used for accessing or modifying the mesh variable data.
	\end{enumerate}
	
	These indexes work in concert to enable fast, reliable direct access to level-set data. 
	For example, to retrieve a data value at specific spatial location, 
	one would first use the cell index to locate the date package and its associated packages index. 
	This is then used to access data in \texttt{MeshVariable}, 
	and finally the data index identifies the specific location within the data package. 
	This design eliminates pointer chasing and enhances compatibility with 
	both CPU and GPU architectures by relying solely on array-based indexing.
	\subsection{Neighbourhood}
	Level set computations, such as interpolation and stencil operations, 
	frequently involve accessing neighboring cell data. 
	In sparse domains, performing a full mesh lookup for neighbors at each access is inefficient. 
	To mitigate this, the previous implementation \cite{yu2023level} 
	employed a larger storage matrix with a skin layer for each data package, 
	storing explicit pointers to the data of neighboring cells in the border regions. 
	While this approach simplified data access during computation, 
	it imposed substantial memory overhead, 
	especially when multiple mesh variables were involved.
	
	The present method introduces 
	a more compact and scalable neighborhood storage model. 
	For each activated cell, a special data-package with subdivision size of 3 is 
	used to indicate the package indexes of the nearest neighboring packages, 
	enabling indirect access without storing physical addresses for each data. 
	When a neighboring data indicated by a shift index is required, 
	an access function \texttt{NeighbourIndexShift}, as shown in Lst. \ref{lst:shift}, 
	computes the target package and data indexes
	based on the logical position on the mesh. 
	This avoids storing direct memory addresses 
	and leverages the high arithmetic throughput of modern processors.
	
	{
		\scriptsize
		\begin{lstlisting}[caption={Indirect data access by index shift}, label={lst:shift}, language=C++]
		template <int PKG_SIZE>
		DataPackagePair NeighbourIndexShift(
		const Arrayi &shift_index, const CellNeighborhood &neighbour)
		{
		DataPackagePair result;
		Arrayi neighbour_index = (shift_index + Arrayi::Constant(PKG_SIZE)) / PKG_SIZE;
		result.first = neighbour(neighbour_index);
		result.second = (shift_index + Arrayi::Constant(PKG_SIZE)) - 
		neighbour_index * PKG_SIZE;
		return result;
		}
		\end{lstlisting} 
	}
	
	Note that above function is only able to access the data in nearest neighbor packages.
	For those data beyond that, a generalized access function is designed 
	for a two step approach by which a cell shift is determined first 
	and then the above access function is applied to finalize the data location.
	\section{Computation Offloading}
	A computational operation acting on a data packages is designed as local mesh dynamics, 
	in respected to the corresponding term, i.e. local particle dynamics, 
	for the operation acting on a SPH particle.
	Similarly to the usage of particle dynamics class for SPH simulation, 
	mesh dynamics class defines the scope of cells over 
	which the local dynamics should be applied.
	While \texttt{ALLMeshDynamics} execute the computing kernel of local mesh dynamics 
	at each cell on the background mesh, 
	\texttt{MeshPackageDynamics} do it on specific type of activated cells, 
	as shown in Lsts. \ref{lst:all_mesh} and \ref{lst:mesh_pkg}.
	
	In this work, SYCL programming model is employed to implement execution of
	computing kernels written in plain \texttt{cpp} 
	without the knowledge of vendor-based GPU programming models, such as CUDA or HIP.
	Since SYCL functions are compatible with both host and device execution-as 
	long as they are defined at compile time—there 
	is no need for separate host and device codes of the kernel. 
	Provided that the data pointers are valid during execution, 
	the same function can be executed seamlessly on the selected SYCL device 
	without modification. 
	This design enables not only device-agnostic execution: 
	once the kernel function is defined and a device is selected, 
	it can be executed anywhere without further changes, 
	but also the ability of debugging and testing device codes 
	in a pure host-side CPU environment. 
	
	{
		\scriptsize
		\begin{lstlisting}[caption={Execution of \texttt{ALLMeshDynamics}}, 
		label={lst:all_mesh}, language=C++]
		UpdateKernel *update_kernel = kernel_implementation_.getComputingKernel();
		mesh_for(
		ExecutionPolicy(), 
		MeshRange(Arrayi::Zero(), index_handler_.AllCells()),
		[&](Arrayi cell_index)
		{
		update_kernel->update(cell_index);
		});
		\end{lstlisting} 
	}

	{
		\scriptsize
		\begin{lstlisting}[caption={Execution of \texttt{MeshPackageDynamics}}, 
		label={lst:mesh_pkg}, language=C++]
		UpdateKernel *update_kernel = kernel_implementation_.getComputingKernel();
		int *pkg_type = dv_pkg_type_.DelegatedData(ExecutionPolicy());
		package_for(
		ExecutionPolicy(), 
		num_singular_pkgs_, sv_num_grid_pkgs_.getValue(),
		[=](UnsignedInt package_index)
		{
		if (pkg_type[package_index] == TYPE)
		update_kernel->update(package_index);
		});
		\end{lstlisting} 
	}
	
	Each mesh dynamics instance contains an associated implementation object, 
	which encapsulates the corresponding operation and 
	is ready for dispatch to the designated execution device. 
	Both the mesh local dynamics class and the execution policy are provided 
	as template parameters to the mesh dynamics, 
	enabling compile-time specialization and lazy dispatch
	for different operation types and target architectures.
	\subsection{Mesh Local Dynamics}
	The class of a specific mesh local dynamics is decomposed into two components: 
	variable management and a computing kernel. 
	Variables involved in an operation must either be stored as copyable small objects 
	or managed by the variables mentioned above. 
	While the small objects are copied directly to the device alongside computing kernel instance, larger data structures are managed via the functions of \texttt{DiscreteVariable}, 
	as will be discussed in the next section, 
	which ensures correct allocation and synchronization across host and device.
	
	Each local dynamics has its own dedicated computing kernel definition 
	as a nested class, ensuring modularity and facilitating kernel reuse. 
	Therefore, every computational operation is thus encapsulated in a class form, 
	enabling the implementation class to construct an executable instance.
	
	During initialization, the mesh local dynamics object is instantiated 
	and configured with all required variables. 
	However, the computing kernel is only initialized at the first request, 
	in which the host-device transfer of data takes place according to the execution policy. 
	Computing is then carried out transparently on the target platform as determined by the policy.
	\subsection{Implementation Class}
	An implementation class is responsible for the computing kernel transfer 
	and provide the mesh dynamics with a valid computing kernel based on execution policy. 
	As shown in Lsts. \ref{lst:all_mesh} and \ref{lst:mesh_pkg}, 
	the computing kernel will be created in the implementation object 
	by a getter function at the first call. 
	According to the execution policy chosen, 
	the implementation object is responsible to copy the computing kernel 
	inside local dynamics instance to the device to execute. 
	As computing kernels take only copyable objects or raw variable pointers, 
	the kernel can be copied to the device directly with no extra attention.
	\subsection{Data Transfer}
	To enable computation on GPU, 
	all relevant data must be resident on the device in advance. 
	In SYCL, two primary memory access mechanisms are available. 
	The first utilizes \texttt{sycl::buffer} and accessors. 
	SYCL automatically manages data transfers between host and device, 
	ensuring that data is available when accessed. 
	This implicit data management simplifies development
	but can reduce control and transparency.
	
	The present work adopts mainly the second unified share memory (USM) mechanism, 
	due to its explicit control and pointer access can 
	used in the same way on host and device with plain \texttt{cpp}. 
	The pointer-based nature of USM simplifies the function interface, 
	as only raw pointers need to be handled. 
	This eliminates the need for buffer management in user-defined functions, 
	allowing them to work directly without aware of the SYCL types and functions.
	
	For data arrays that requires updates, 
	such as mesh, background-mesh and meta variables, 
	they are managed by the \texttt{DiscreteVariable} class. 
	Each \texttt{DiscreteVariable} includes 
	a \texttt{DeviceOnlyVariable} instance that 
	manages the device-side allocation, transfer, synchronization, and deallocation, ensuring correct memory management and preventing leaks.
	Note that, device-side memory is only allocated 
	and data transferred when the device pointer is first requested.
	Again the latter is generated only when a computing kernel is first requested by 
	the loop function \texttt{mesh\_for} \texttt{package\_for},
	as shown in Lst. \ref{lst:all_mesh} and \ref{lst:mesh_pkg}, 
	with device execution policies.
	
	For singular global parameters, such as total number of data packages, 
	memory is allocated using \texttt{malloc\_shared} from the USM mechanism.
	This shared memory is accessible from both host and device, 
	with synchronization handled automatically by SYCL.
	\section{Multi-resolution Level-set}
	The level set is constructed from an input geometry (STL file) 
	using a multi-resolution approach in order to ensure water-tight surface 
	when the input geometries are leaking due to the topological inconsistency. 
	Given a target resolution, 
	several layers with successively doubled resolutions are established. 
	Each layer corresponds to a distinct spatial resolution 
	as an instance of \texttt{MeshWithGridDataPackages}, 
	which encapsulates both level-set data and metadata for that layer. 
	The coarsest mesh is initialized first, 
	and subsequent layers that each with doubled resolution are initialized 
	based on the preceding coarser layer. 
	The geometric shape and boundary conditions remain consistent across all layers; 
	but the finer layer contains 4 (2D) or 8 (3D) times as many logical cells 
	as its coarser counterpart. 
	This successive initialization continues until the desired resolution is achieved. 
	The number of required layers is computed in advance.
	
	\subsection{Initialization Process}
	The present initialization process is mainly carried out on CPUs
	due to the present signed-distance function from a triangle mesh 
	\cite{baerentzen2005robust} lacking of GPU implementation.
	Specifically, the initialization process is composed of the following steps.

	\begin{enumerate}
		\item Initial cell tagging for the core data packages. 
		For the coarsest layer, 
		the distance from each cell of the background mesh to the surface is evaluated 
		and whose distances are less than the cell size is activated.
		For successive layers, 
		only the cells covered by the coarse core cells will be evaluated. 
		Using Intel's TBB (Threading building Blocks) concurrent vector, 
		the cell index and packages type are inserted.
		\item Tagging the cell for inner data packages. This is done by checking if any nearest neighbor of a mesh cell activated by a core data package. Again, the cell index and packages type are inserted to the concurrent vector.
		\item Sorting for locality. During tagging, 
		activated cells are identified in parallel. 
		This results that the indexes of activated cells are in arbitrary order, 
		which may degrade cache locality. To address this, a sorting step is introduced to reorder cell indexes according a predefined sequence, 
		typically based on spatial location. After sorting, 
		the cell indexes and data-package types are then copied 
		to newly allocated meta variables.
		So that they can be used later in GPU computing.
		\item After that the level-set value is evaluated for all data packages using the sign-distance function from triangle mesh \cite{baerentzen2005robust}.
		\item Finally, the neighborhood of all activated cells are defined. 
		Note that, in order to achieve full consistency for data queries, 
		the far-field packages' neighbors are set to themselves. 
	\end{enumerate}
	
	Note that, after the level-set value is evaluated for all data packages, 
	the computation can be carried out either on CPU or GPU 
	according ro the chosen execution policies.
	Also note that, the level-set value evaluation can be carried before the sorting, which then can be run on GPU but need to reorder 
	the level-set data packages evaluated in arbitrary order.
	\subsection{Consistency Correction}
	Due to the possible topological inconsistency of STL file, 
	directly application of signed-distance function \cite{baerentzen2005robust} 
	to identify the contain condition (or the sign) of level set may go wrong.
	Such issue may finally leads to the generation of SPH particles wrong. 
	To avoid this issue, only the sign of level set for those data points
	very close to the surface is directly used,
	those at other locations are obtained by a two-step diffusion process from the near interface to the entire domain. The first coarse step is on the mesh cells and the second refined one is on the data packages. Note that only on the coarsest layer all cells are evaluated. For refined layers, the operation is limited to inner cells or data packages. 
	\subsection{Small Feature Cleaning}
	Very often the geometry which is originally generated for manufacturing 
	includes many small features which are not necessary for computational fluid or solid dynamics (CFD or CSD) simulations 
	and may lead to numerical instabilities if not cleaned.
	In the present work, we reimplemented the level-set cleaning algorithms 
	(only on the finest layer) in Ref. \cite{yu2023level} so that it can be run on GPU.
	
	Note that, these algorithms heavily 
	replies on the indirect access to neighbor data packages.
	As shown in Lst. \ref{lst:stencil}, a typical regularized central difference scheme 
	for gradient evaluation can be 
	implemented with the help of \texttt{NeighbourIndexShift}. 
	Also note that efficient high-order finite difference can be extended
	easily due to the relative large size of the data package.
	
	{
		\scriptsize
		\begin{lstlisting}[caption={A 2D regularized central differnece scheme}, 
		label={lst:stencil}, language=C++]
		template <int PKG_SIZE, typename RegularizeFunction>
		Vec2d regularizedCentralDifference(
		PackageData<Real, PKG_SIZE> *input, const CellNeighborhood2d &neighborhood,
		const Array2i &data_index, const RegularizeFunction &regularize_function)
		{
		DataPackagePair center = 
		NeighbourIndexShift<PKG_SIZE>(data_index, neighborhood);
		DataPackagePair x1 = NeighbourIndexShift<PKG_SIZE>(
		data_index + Array2i(1, 0), neighborhood);
		DataPackagePair x2 = NeighbourIndexShift<PKG_SIZE>(
		data_index + Array2i(-1, 0), neighborhood);
		DataPackagePair y1 = NeighbourIndexShift<PKG_SIZE>(
		data_index + Array2i(0, 1), neighborhood);
		DataPackagePair y2 = NeighbourIndexShift<PKG_SIZE>(
		data_index + Array2i(0, -1), neighborhood);
		Real dphidx_p = input[x1.first](x1.second) - input[center.first](center.second);
		Real dphidx_m = input[center.first](center.second) - input[x2.first](x2.second);
		Real dphidx = regularize_function(dphidx_p, dphidx_m);
		Real dphidy_p = input[y1.first](y1.second) - input[center.first](center.second);
		Real dphidy_m = input[center.first](center.second) - input[y2.first](y2.second);
		Real dphidy = regularize_function(dphidy_p, dphidy_m);
		return Vec2d(dphidx, dphidy);
		}
		\end{lstlisting} 
	}
	\section{Integral with SPH Smoothing Kernel and Grid-Particle Coupling}
	Another algorithm in Ref. \cite{yu2023level} is computing 
	the integral with SPH smoothing kernel on the sparse-grid. 
	This is essential for the physical relaxation of the SPH particles \cite{zhu2021cad},
	which is also reimplemented for GPU execution. 
	The SPH particle relaxation is a typical grid-particle coupling algorithm 
	in which the integral field is interpolated
	by bi- or tri-linear interpolation to the particle's position 
	and used in the form of surface force to drive the particle.
	Note that, the interpolation operation is different from the previous local mesh dynamics 
	as it, other than looping on the cells or data packages, 
	but evolves position-based random memory access of a data package and may be its neighbors,
	which again heavily relies on the usage of \texttt{NeighbourIndexShift}. 
	\section{Performance Evaluation}
	Compared to the previous implementation on SPHinXsys, 
	memory usage has been significantly reduced, in the present work, 
	by eliminating the need to store a data address 
	for each individual data entry within an activated cell. 
	Instead, a \texttt{size\_t}-typed package index, 
	as shown in Lst. \ref{lst:mesh_class}, 
	is now used to retrieve neighboring package indexes. 
	As a result, memory usage is reduced by $6\times 6 \times 8=288$ bytes per cell per mesh variable in each layer. 
	Given the limited memory resources on GPU devices, 
	this optimization is particularly valuable.
	
	\begin{table*}[htbp]
		\centering
		\begin{tabular}{|c|c|c|c|}
			\hline
			& OpenVDB & SPGrid & SPHinXsys \\
			\hline
			Sequential, 1 thread  & 79.563 & 77.2598 & 22.948 \\
			Sequential, 4 threads & 32.322 & 29.8752 & 7.429 \\
			Stencil, 1 thread     & 1013.162 & 229.572 & 59.972 \\
			Stencil, 4 threads    & 303.902 & 68.8437 & 21.378 \\
			\hline
		\end{tabular}
		\caption{Timing for operations for SPGrid, OpenVDB, and SPHinXsys.}
		\label{tab:example}
	\end{table*}

	To evaluate the computational performance 
	with previous popular sparse-grid methods, namely OpenVDB and SPGrid, 
	a shelled sphere is used as the shared test geometry across all methods. 
	The shell is centered at (0.5, 0.5, 0.5) and has an inner radius of 0.3 and outer one of 0.31, with a resolution of 1/1024.
	A sequential access and a stencil operation are carried out on all the activated data.
	
	Note that, for sequential access, the \texttt{LeafManager} is employed for OpenVDB. 
	It is initialized with the target grid, after which each leaf node is iterated. Within each leaf, the offset filter provided by OpenVDB is used to iterate through active data points. 
	The sequential benchmark involves making minor changes to each value 
	stored in an activated cell.
	
	For stencil operations, the seven-point Laplacian operator provided by OpenVDB is used. 
	A comparable stencil computation is implemented 
	by using \texttt{NeighbourIndexShift} for SPHinXsys. 
	In both cases, all active data are visited, 
	and a Laplacian operation is applied at each active data too.
	Note that, due to the lack of reference on GPU execution,
	here, we only provide the CPU performance of the present method.  
	Also note that, for this test OpenVDB requires only one-third of the data storage compared to SPHinXsys. This difference is expected, as OpenVDB initializes level sets on a data-wise basis, allowing each data to be individually categorized.
	On the other hand, when data access and stecil operations are tested, 
	relative more data will be accessed or modified in the present method.
	
	The evaluation results for both single and multi-thread execution 
	are shown in Tab. \ref{tab:example},
	which shows, in both scenarios, the presented method outperforms OpenVDB and SPGrid. 
	This improvement is attributed to the overhead of tree traversal required during neighbor lookup in their methods. Although OpenVDB implements a software cache to mitigate this issue, the traversal process remains a significant bottleneck, especially when frequent neighbor access is required for operations like the Laplacian.
	In addition, the prefetch enabled by sequential access of contiguous memory storage 
	also significantly influence the results.
	\section{Numerical Examples}
	Here, we demonstrate two typical applications, 
	both as pre-processing tool for SPH simulation, 
	of the present method.
	
\begin{figure}[htb!]
		\centering
		\includegraphics[width=\linewidth]{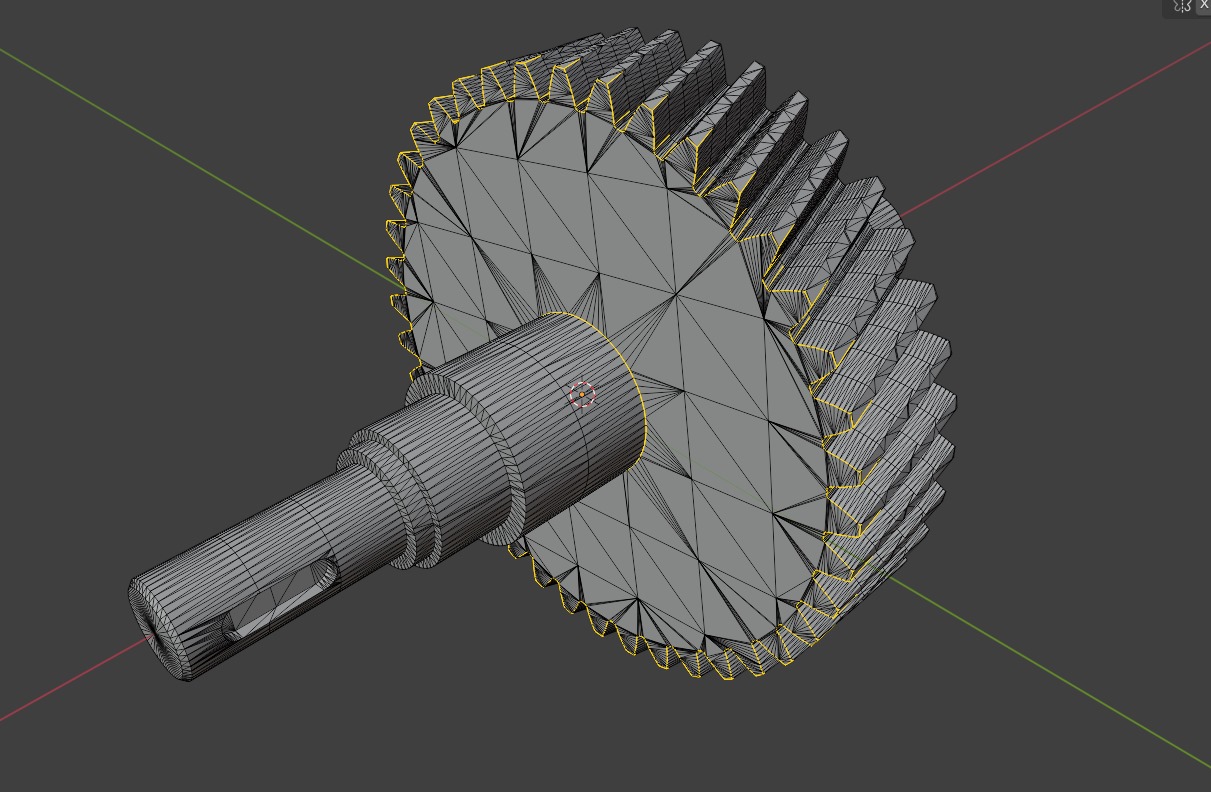}
		\caption{STL triangle mesh of a gear. The leaking regions are indicated with yellow lines.}
		\label{fig:stl-gear}
	\end{figure}

\begin{figure}[htb!]
		\centering
		\includegraphics[width=\linewidth]{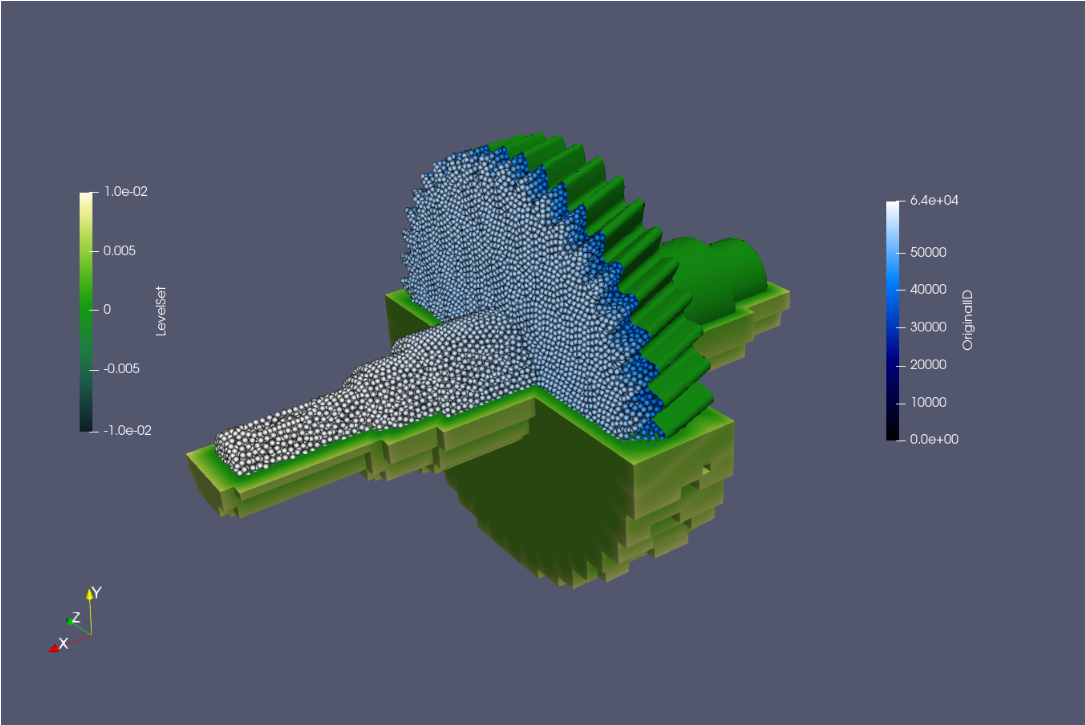}
		\caption{Corrected level-set field, zero-level contour and the finally generated SPH particles. The zero-level contour is colored with green.}
		\label{fig:gear-particles}
	\end{figure}

	In the first case, SPH particles are generated from a gear model in STL file 
	for a gearbox simulation, as shown in Fig. \ref{fig:stl-gear}. 
	While the STL model visually is perfectly water-tight, 
	there are many triangle faces are not strictly connected and a straightforward application of the signed-distance function from a triangle mesh 
	\cite{baerentzen2005robust} will leads to many particles generated (leaked) 
	away from the surface due to the contain condition is wrongly predicted.
	With the present method, after the initial level set is generated, 
	the contain consistency correction algorithm is applied and run on GPU 
	with high computational efficiency.
	The corrected level set field and final generated SPH particles 
	is shown in Fig. \ref{fig:gear-particles}. 
	Note that gear surface is represented with a narrow band of (core) data packages, 
	which is able to achieve memory efficiency 
	and is able to handle those level-set based 
	algorithms evolving data locations considerably far from the surface.
	
\begin{figure}[htb!]
		\centering
		\includegraphics[width=\linewidth]{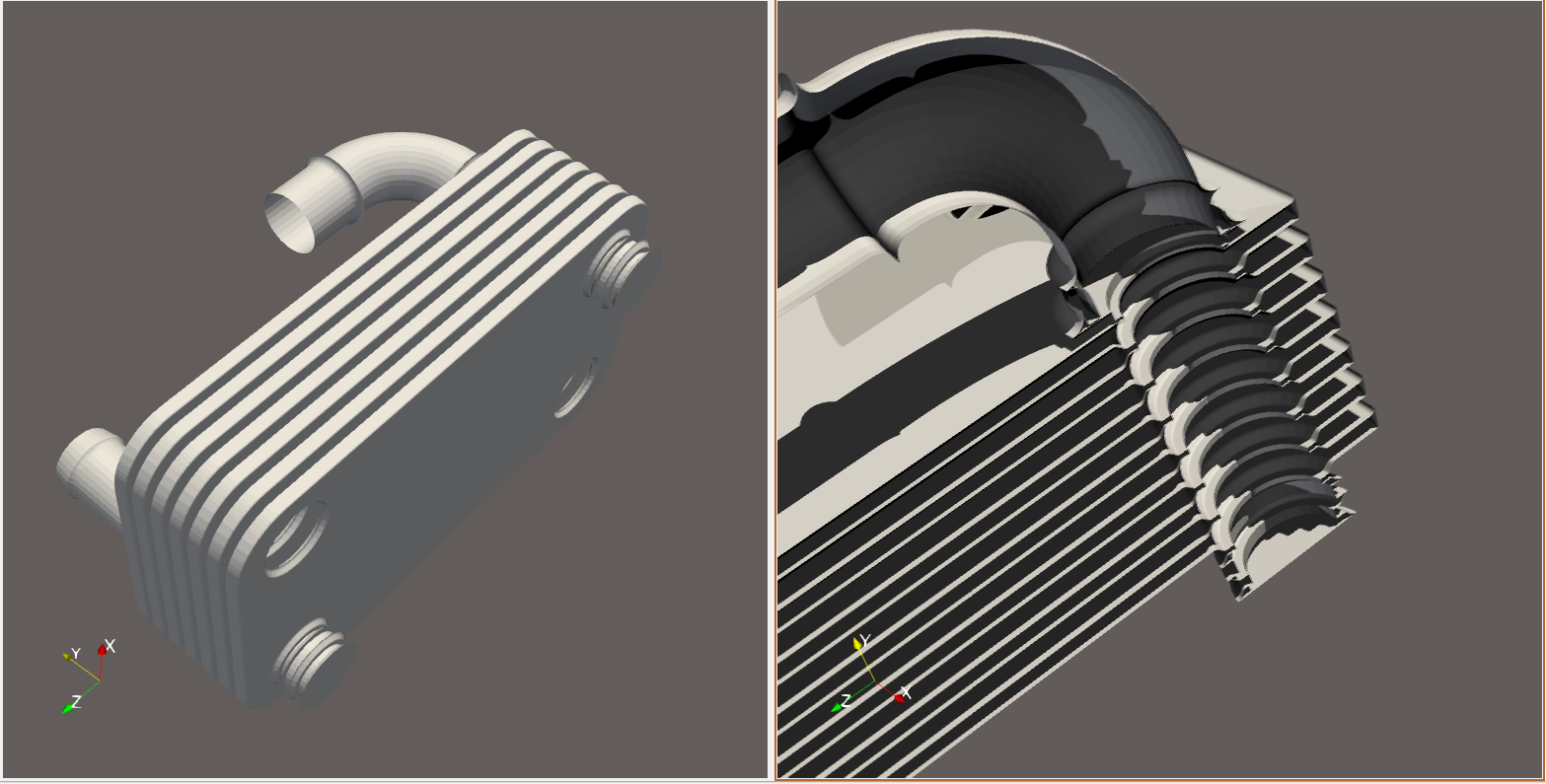}
		\caption{Geometric model of a heat exchanger. (left) the entire model. 
			(right) a cross-section view indicting very small features. }
		\label{fig:heat-exchanger}
	\end{figure}

\begin{figure}[htb!]
		\centering
		\includegraphics[width=\linewidth]{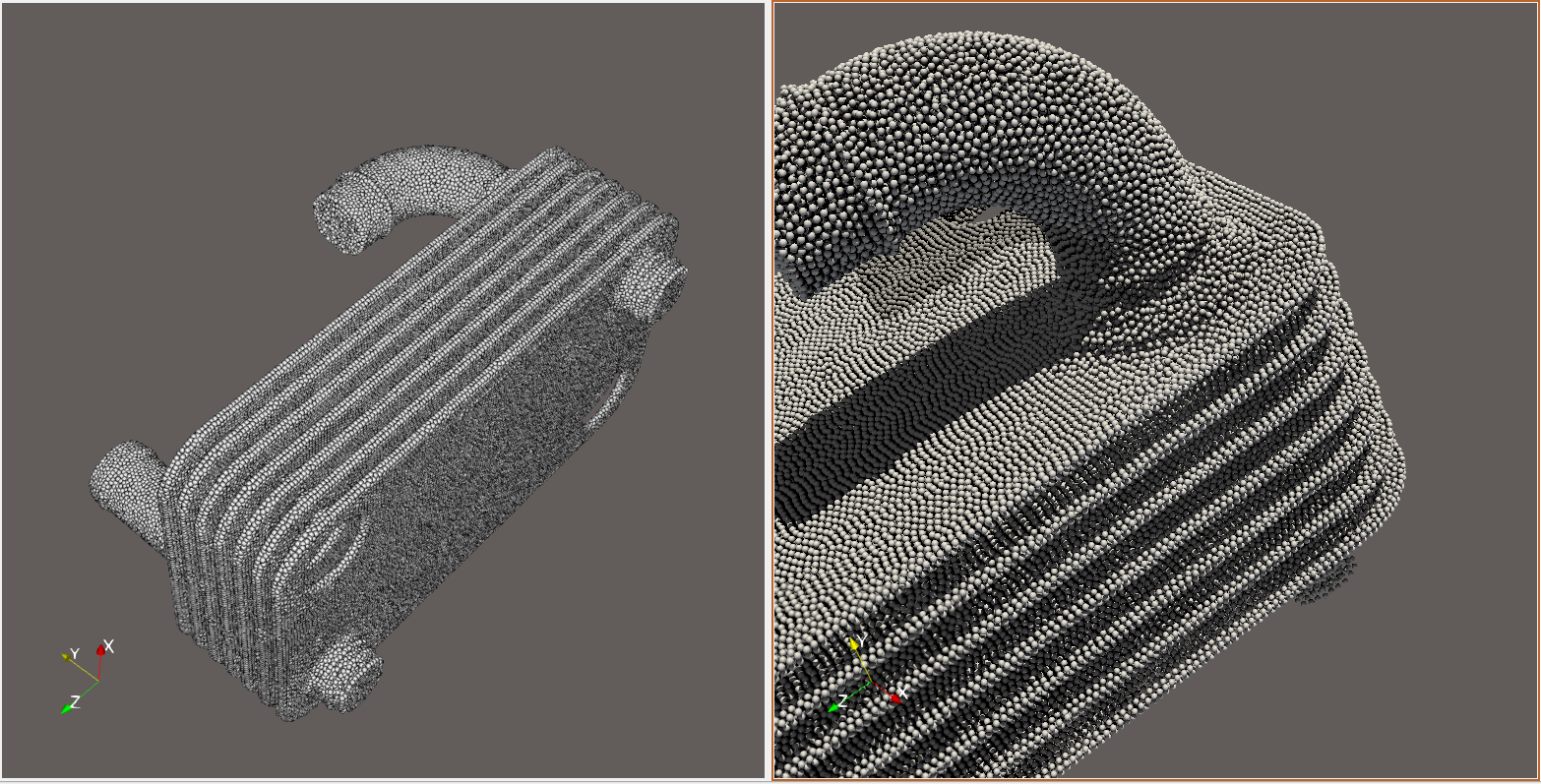}
		\caption{SPH particles for a heat exchanger model. (left) the entire model. 
			(right) a detail-view near the region where small features are cleaned. }
		\label{fig:heat-exchanger_particle}
	\end{figure}

	Second example is to handle a industrial design of heat exchanger, 
	as shown in Fig. \ref{fig:heat-exchanger}, 
	where very small geometry features are presented near 
	the connections between pipes and plates. 
	Note that, while these features can be captured accuracy using 
	high-resolution level-set field, they are not necessarily 
	for a optimization-oriented numerical simulations 
	focusing on the heat-transfer performance only.
	As shown in Fig. \ref{fig:heat-exchanger_particle}, 
	these small features are removed by applying the cleaning algorithm \cite{yu2023level}.
	With the present method, this operation can be carried out on GPU, 
	and decreases computation time greatly (with about 5 times speed up).
	As shown in the finally generated SPH particles,  
	a regular particle distribution without single layer or 
	filament of particles generated and hence ensure stable SPH simulations. 
	\section{Limitations and Future Work}
	It is noteworthy that, compared previous sparse-grid method, 
	The present approach marks an entire data package as activated if its center overlaps with the body surface, regardless of the actual volume intersected. Such tagging process can be further
	optimized to reduce memory redundancy while maintaining computational accuracy.
	Another future work would be developing GPU-enabled 
	sign-distance-from-triangle-mesh algorithm,
	so that the full sparse-grid method can be run on GPU 
	for maximum computational efficiency.
	\bibliographystyle{ACM-Reference-Format}
	\bibliography{xyhu-base}
\end{document}